\documentstyle[aps,preprint]{revtex}

\def\squote{}
\def\quote#1#2#3#4{\squote {#1,\ {\sl#2}\ {\bf#3}, #4}.\par} 
\def\qquote#1#2#3#4{\squote {#1,\ {\sl#2}\ {\bf#3}, #4};} 
\def\nquote#1#2#3#4{\squote {#1,\ {\sl#2}\ {\bf#3}, #4}}

\def\prl{{\sl Phys. Rev. Lett.}\ }

\def\pr {{\sl Phys. Rev.}\ }
\def\ksi{\xi}

\def\ES{Efros-Shklovskii\ }
\def\dos{density of states\ }

\def\sss                               {
     \scriptscriptstyle                       }

\def\e{\epsilon}

\begin{document}
\draft

\title{Universal Crossover between Efros-Shklovskii and Mott 
Variable-Range-Hopping Regimes}
\bigskip
\author{Yigal Meir}
\address{Physics Department, Ben Gurion University, Beer Sheva 84105, ISRAEL\\
and\\Institute of Theoretical Physics, University of California at Santa Barbara,
\\Santa Barbara, CA 93106}
\bigskip
\maketitle
\begin{abstract}
A universal scaling function, describing the crossover between the Mott
 and the Efros-Shklovskii hopping regimes, is derived,
using the percolation picture of transport in strongly localized systems.
 This function is agrees very well
with experimental data. Quantitative comparison with experiment allows for
 the possible determination of the role played by polarons in the transport.
\end{abstract}
Electronic interactions are known to play an important role in the strongly
localized regime. More than two decades ago, Pollak \cite{pollak}, and
Efros and Shklovskii \cite{es}
 pointed out that the long-range nature of the interactions leads to
a dip in the single-particle density of states,  $\rho(\e)$,  at the Fermi 
energy.  Using a constraint on the single-particle excitations, 
Efros and Shklovskii \cite{es} argued that this soft gap is of the 
form $\rho(\e)\sim \e^{d-1}$,  where $\e$ is the energy measured from the 
Fermi energy and $d$ the space dimension. The gap in the density of states
was indeed observed by tunneling \cite{mcmillan,mlee} and photoemission
\cite{photo} measurements. Moreover, it 
 modifies the Mott variable-range hopping law,   $\log R \sim 1/T^x$, 
from $x=d+1$ to $x=1/2$. The modified exponent has been observed in many
experiments \cite{esbook}.

The \ES hopping law is expected to be relevant at low temperatures (compared
to the size of the gap) \cite{es,ora},  and in the last few years there have
been a multitude of experiments aimed at exploring the crossover between the
Mott to the \ES hopping regimes as a function of temperature \cite{crossover}.
 A significant step was taken more recently, when 
Aharony and coworkers \cite{amnon,lien} have argued that this crossover is
described by a universal scaling function,  and obtained this scaling
function phenomenologically,  using energy additivity. 
It is the aim of this paper to demonstrate that the
``microscopic" percolation picture \cite{ambegaokar}, 
 describing the transport in the strongly
localized regime, indeed leads to such a universal crossover
function,  and to derive an equation for this function. The results agree
excellently with existing experimental data \cite{mlee},
 and allow quantitative 
comparison between features in the density of the states and the transport
data, taken on the same physical system.

The starting point of this calculation is the mapping of the resistance 
 problem into a percolation criterion \cite{ambegaokar} 
 of an equivalent
 random resistor network \cite{miller}, consisting of randomly placed
 sites. Without interactions,  the activation energy from site $i$ to site $j$
 is given by $\e_j-\e_i$,  where $\e_i$ are the energy of site $i$,  which
 in this case is distributed uniformly.
The resistance 
 between each pair of sites is given by \cite{miller} 
$R_{ij} = \exp\left\{(|\e_i|+|\e_j|+|\e_i-\e_j|)|/2T + 2 r_{ij}/\ksi 
\right\}$,  where all resistances are measured in some unit
of resistance.  
Since the resistances vary exponentially, the overall
 resistance will be determined by the weakest link, which is the largest
 resistance, $R$, such that 
 the cluster formed by all resistances (bonds), satisfying $R>R_{ij}$,
percolates. Clearly, all states  participating in the percolating
network (defined as occupied sites) must satisfy  $R>\exp(|\e_i|/2T)$.
 Following \cite{ambegaokar}, the percolation criterion employed here 
is the following \cite{spinvrh}
 --- given such an occupied site , the number of bonds
 attached to it has to be higher than a critical threshold, $Z_c$,
 for the system to percolate,
\begin{eqnarray}
 Z_c &=& {1\over {2 \int_{\sss  -T\log R}^{\sss T\log R}\rho_0 d\e}} \int_{-2T\log R}^{2T\log R}
 \rho_0\, d\e_1\  \int \rho_0\, d\e_2\ 
 \int  d^d r_{12}\  \Theta(R - e^{(|\e_1|+|\e_2|+|\e_1-\e_2|)/2T +
 2 r_{12}/\ksi}) \nonumber\\
&=& 3 T {{\ksi^d\rho_0\pi^{d/2}(\log R)^{d+1}}\over{2^{d+1}\Gamma(d/2)d(d+1)(d+2)}} , 
 \nonumber\\
&\equiv& Z_c T(\log R)^{d+1}/T_0  , 
\label{pc}
\end{eqnarray}
leading directly to the Mott hopping low , $\log R = (T_0/T)^{1/(d+1)}$ 
(where the last equality defines $T_0$).
In the above $\rho_0$ is the uniform density of states, and $\ksi$ the 
localization length.

In the presence of interactions, the activation energy 
from site $i$ to site $j$ is given by\cite{es} $\e_j-\e_i-e^2/r_{ij}$, 
 where $r_{ij}$ is the distance between the sites,  
 (where all other particles are
 assumed to remain fixed). Since an activation
 from the ground state has to be positive,  this leads to a constraint
 on the distribution of $\e_i$,  and to the soft gap in the Fermi 
surface\cite{es}.
 In fact,  given that constraint,  Efros \cite{efros,raikh} has been able to 
 derive  a self-consistent equation for the density of states, 
\begin{equation} 
\rho(\e)/\rho_0 = \exp \left\{ 
\alpha \int_0^\infty d\e_1 {\rho(\e_1)\over{(\e+\e_1)^d}} \right\}.
\label{dos}
\end{equation}  
where $\alpha$ is some numerical constant, and 
an infinite band was assumed. 
The resulting \dos is clearly of the form $\rho_0 f(\e/\e_0)$,  with
 $\e_0\sim (e^{2d}\rho_0)^{1/(d-1)}$,  
and it can be very well approximated in three dimensions
 by $f(x)= x^2/(1+x^2)$, with $\e_0\simeq0.46 e^3\sqrt{\rho_0}$.

Knotek and Pollak \cite{knotek} and Mott \cite{mott} have pointed out 
that the resistance is not determined by the single-electron density of 
states, but should involve many electron excitations (polarons). 
Efros \cite{efros} indeed demonstrated that taking into account 
multi-particle excitations (i.e. not assuming that all other particles remain
fixed, when electron is activated from site $i$ to site $j$),
  leads to an even stronger (exponential) 
gap in the \dos \cite{efros,parabolic}. The long-range interactions
between the polarons, which are indeed the relevant excitations for the
resistance,  lead to a density of states which satisfies (\ref{dos}), and
consequently the resistance should be of the Efros-Shklovskii type,
$\log R = (T_{ES}/T)^{1/2}$.

 Taking these considerations into account, 
the percolation condition (Eq.(\ref{pc})) is now modified in two ways. The
probability of finding two sites,  of energies $\e_i$ and $\e_j$ and 
distance $r_{ij}$ apart is proportional to
 $\rho(\e_1)\rho(\e_2)\Theta(\e_2-\e_1-e^2/r_{12})$, 
 where $\rho(\e)$ satisfies (\ref{dos}),  and the activation energy 
 from site $i$ to site $j$ is given by
$\e_j-\e_i-e^2/r_{ij}$. The resulting percolation condition
 can now be written in the form, 
 \begin{eqnarray}
 Z_c &=& {1\over {\int_{\sss -T\log R}^0 \rho(\e) d\e}}
 \int_{-2T\log R}^0 \rho(\e_1)\, d\e_1\ 
 \int_0^{2T\log R} \rho(\e_2)\,  d\e_2\ 
 \nonumber\\ 
 & & \int  d^d r_{12}\  \Theta(\e_2-\e_1-\e^2/r_{12}) 
 \, \Theta(R - e^{(\e_2-\e_1-e^2/r_{12})/T + 2r_{12}/\ksi})
\label{pcinter}
\end{eqnarray}
Using the one-parameter scaling form of $\rho(\e)$,  and defining
the dimensionless parameters $x$ and $y$ by
$1/T= \alpha_2 (e^2/\ksi) x /\e_0^2$ and 
$\log R = \alpha_1 (e^2/\ksi) y/\e_0$,
where $\alpha_i$ are numerical amplitudes,  this
equation can be recast in the form
\begin{eqnarray} 
1 = \int_{-(y'+s_1)/2x'}^0 f(\e_1)\,d\e_1 
\int_{\max\{\e_1+(y'-s_1)/2x', 0\}}^{\e_1+(y'+s_1)/2x'}
& & f(\e_2)\,d\e_2\left\{(s_2+\sqrt{s_2^{^2}+8x'})^d/4^d-1/(\e_2-\e_1)^d
\right\}  
\nonumber\\ &{\large /}& \ \ \int_{(-y'+s_1)/2x'}^0 f(\e_1), 
\label{Fscaling}
\end{eqnarray}
with $x'=\alpha_2 x$, $y'=\alpha_1 y$,  $s_1=\sqrt{y'^2-16x'}$,  and
$s_2=y'-(\e_2-\e_1)x$.
Eq.(\ref{Fscaling}) which is an implicit equation for $y$ in terms of $x$, 
or $\log R$ in terms of $T$,  
indeed demonstrates that the resistance can be written
in a scaling form
\begin{equation} 
\log R =  A f\left(T/T_x\right) , 
\label{Rscaling}
\end{equation}
in agreement with Aharony et al. \cite{amnon}, 
with $A\equiv(T_0/\e_0)^{1/d}$,  and $T_x\equiv(\e_0^{d+1}/T_0)^{1/d}$ (the
numbers $\alpha_i$ were chosen such that the amplitudes of $1/T^x$ in the
two asymptotic regimes of $f(x)$
will agree with the amplitudes chosen by Aharony et al.). 
 In addition to relating the nonuniversal amplitudes $T_x$ and
$A$ to $\e_0$,  namely
to properties of the \dos, which can be independently measured, 
 Eq.(\ref{Fscaling}) gives a prescription to calculating the full scaling 
 function from ``first principles''.
 
We have calculated the
crossover function in three dimensions,  
using Eq.(\ref{Fscaling}) and a density of states of the quasi-particles
of the form $\rho(\e)=\rho_0\e^2/(\e^2+\e_0^2)$.
 In Fig.~1 we compare the derivative of the scaling function to the data
published in Ref.\cite{mlee}. In comparison we also plot the phenomenological 
function derived by Aharony et al. \cite{amnon}. Two points are noteworthy -
 the experimental data exhibits a rather fast crossover from the Mott regime 
 ($\log R \sim 1/T^{1/4}$) to the Efros-Shklovskii regime 
($\log R \sim  1/T^{1/2}$).
The presently derived scaling function
shows an excellent agreement with the data.
The phenomenological scaling function,  and also the scaling function
derived from the procedure suggested by Lien \cite{lien},  show a rather
smooth crossover,  which has a poorer agreement with the data
(see inset). Moreover, as one crosses over to the Efros-Shklovskii regime, 
the scaling function (\ref{Fscaling}) predicts that instead of a monotonic
increase in the exponent from $1/4$ to $1/2$,  there is an overshoot in the
slope,  a feature which is clearly seen in the experimental data. 

There are two physical parameters which can be extracted from the fit ---
$T_0$ and $\e_0$. The value of $T_0$ was found to be $1500$ Kelvins,  exactly
as deduced in the experimental paper from the high temperature data. More
interestingly,  the best fitting value of $\e_0$ was $0.04$ meV, an
order of magnitude smaller than what one would deduce from the single-electron
density of states,  measured {\sl on the same sample},  by tunneling 
spectroscopy.  This seems to be the first verification of the fact that it
is not the single-electron density of states that determines the resistivity, 
but rather the density of states of the dressed particles,  the polarons,  which
has a gap which is much smaller,  due to the finite size of the polaronic
cloud. The comparison between the present calculation and the experiment allows
for the determination of the average cloud size,  which is a few atomic lengths. 

To conclude,  we have used the percolation picture to derive the crossover
function from the Mott to the Efros-Shklovskii hopping regime. This function
shows an excellent agreement with experimental data. Quantitative
comparison with the experiment allows a quantitative analysis
of the effect played by the many-particle excitations in the system. We hope
that this work will stimulate more work in this direction,  especially 
experimental investigations of the resistivity and the density of states
in the same samples, such as those pioneered by Massey and Lee \cite{mlee}.

{\sl Acknowledgments:} I thank Ora Entin-Wohlman for motivating me to work on
this subject,  to Y. Gefen and L. Levitov for fruitful discussions and to
Mark Lee for making the experimental data of 
 Ref. \cite{mlee} available.
 Work at Santa Barbara was supported in part by the {\sl National
 Science Foundation} under Grant No. PHY94-07194.


\vskip 6truecm
\leftline{\sl Figure Caption}

Comparison of the experimental data (diamonds) to the derivative 
of the scaling function derived
in this work (continuous line),  and that of Ref. \cite{amnon} (broken line). 
The data shows
a fast crossover,  and is fitted very well by the function derived here. The
inset shows a comparison with the previously derived functions [Ref. 
\cite{amnon} (broken line),  and Ref. \cite{lien} (dotted line)],  both 
showing too smooth a crossover,  compared to the data.

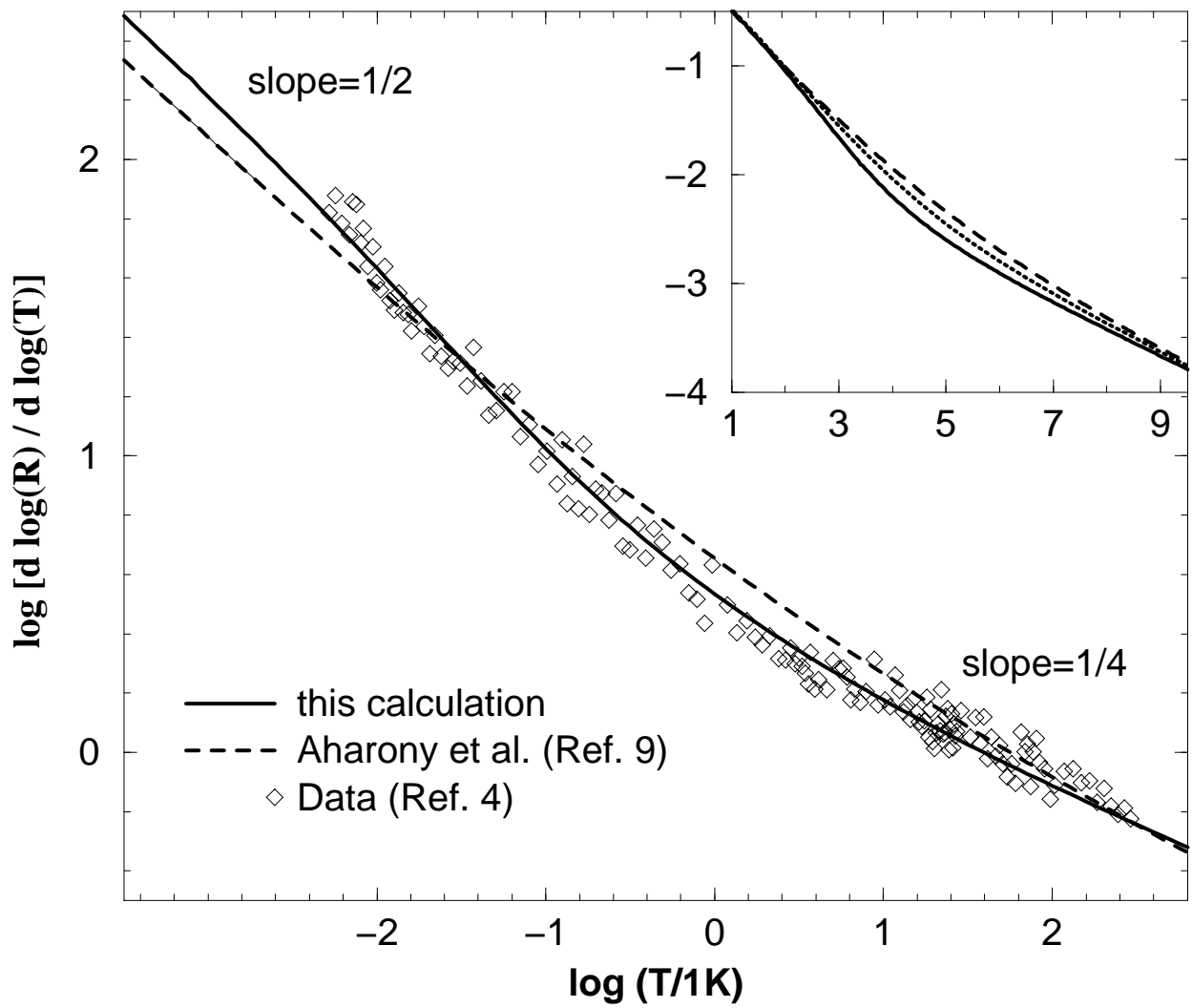
\end{document}